\documentclass{desyproc}

\begin{document}
\title{The Klash Proposal: Status and Perspectives}

\author{{\slshape
   C. Gatti$^1$, D. Alesini$^1$, D. Babusci$^1$,C. Braggio$^{6,7}$, G. Carugno$^{6,7}$,\\
    N. Crescini$^{5,7}$, D. Di Gioacchino$^1$, P. Falferi$^{3,4}$, G. Lamanna$^2$, C. Ligi$^1$,\\
    A. Ortolan$^5$, L. Pellegrino$^1$, A. Rettaroli$^1$, G. Ruoso$^5$, S.Tocci$^1$}\\[1ex]
$^1$ INFN, Laboratori Nazionali di Frascati, Italy\\
$^2$ Universit\`a di Pisa and INFN, Italy\\
$^3$Istituto di Fotonica e Nanotecnologie, CNR—Fondazione Bruno Kessler, Povo, Trento, Italy\\
$^4$INFN, TIFPA,  Povo (TN), Italy\\
$^5$INFN, Laboratori Nazionali di Legnaro,  Legnaro (PD), Italy\\
$^6$INFN, Sezione di Padova, Padova, Italy\\
$^7$Dip. di Fisica e Astronomia,  Padova, Italy\\
}
\contribID{Gatti\_Claudio}

\confID{20012}  
\desyproc{DESY-PROC-2018-03}
\acronym{Patras 2018} 
\doi  

\maketitle

\begin{abstract}
  Recently some of the authors proposed~\cite{Klash:arxiv} a search for galactic axions with mass about 0.2~$\mu$eV using a
  large volume resonant cavity, tens of cubic meters, cooled down to 4~K and immersed in a
  magnetic field of about 0.6~T generated inside the superconducting magnet of the
  KLOE experiment located at the National Laboratory of Frascati of INFN.
  This experiment, called KLASH (KLoe magnet for Axion SearcH), has a
  potential sensitivity on the axion-to-photon coupling, $g_{a\gamma\gamma}$, of
  about $6\times10^{-17}$ $\mbox{GeV}^{-1}$, reaching the region predicted by
  KSVZ~\cite{KSVZ} and DFSZ~\cite{DFSZ} models of QCD axions. We report here the status of the project.
\end{abstract}

\section{Introduction}
The search for galactic-axion with a "Haloscope'', a resonant cavity immersed in a strong magnetic field as proposed by P.~Sikivie~\cite{Sikivie}, is a well established technique. The power generated by axion conversion in a haloscope is proportional to the product of the stored magnetic energy, the loaded quality factor, $Q_L$, and the cavity angular frequency $\omega_{c}$: $P_{\mbox{sig}}\propto \omega_{c}B^2VQ_{L}$. Since cavity volume, $V$, scales approximately as $\omega_c^{-3}$ and the quality factor for a copper cavity at cryogenic temperatures scales as $\omega^{-2/3}$, we have $P_{\mbox{sig}}\propto \omega_{c}^{-8/3}$ making the research at lower masses advantageous even with a moderate magnetic field $B$. ADMX~\cite{ADMX} probed the QCD-axions in the mass range between 2 and 3 $\mu$eV by using a haloscope composed of a resonant cavity with volume 0.2~m$^3$ and unloaded quality factor about $2\times 10^5$ immersed in a magnetic field of 7.6~T. Lower masses are foreseen, for instance, by models where the Peccey-Quinn symmetry is broken before Inflation. In the following, we show that a search for galactic axions with mass in the range 0.3-1~$\mu$eV is possible using a large volume resonant cavity of about $33~\mbox{m}^3$ cooled down to 4.5~K and immersed in a magnetic field of 0.6~T generated inside the superconducting magnet of the KLOE experiment located at the National Laboratory of Frascati of INFN.

\section{The KLOE Superconducting Magnet}\label{sec:magnet}
The KLOE experiment~\cite{KLOEreview} recorded data from $e^{+} e^{-}$ collisions at DA$\Phi$NE, the $\phi$-factory at the Laboratori Nazionali di Frascati (LNF), since April 1999. The detector is surrounded by a NbTi Superconducting magnet~\cite{KLOEMAG,MODENA} providing an homogeneus 0.6 T field. The magnet cryostat has an inner diameter of 4.86~m and a length of 4.4~m. The coil is kept at a temperature of 4.5~K with liquid helium. The helium flow comes from a cryogenic plant with an extra cooling capacity which has been formerly dedicated to the FINUDA experiment and to the four compensating solenoids operating in the DAFNE collider.

\section{The KLASH Haloscope}\label{sec:klash}

\begin{figure}[htbp]
  \begin{center}
    \includegraphics[totalheight=5cm]{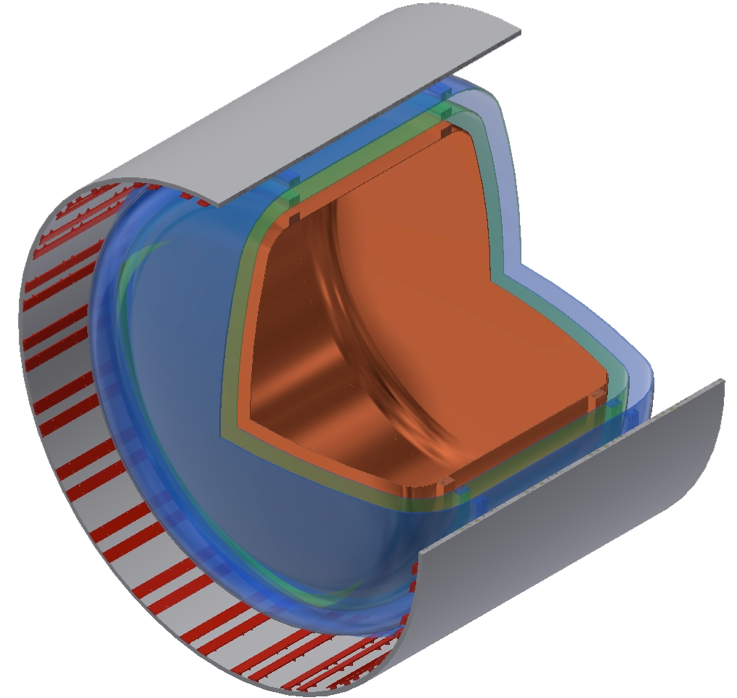}\hspace*{2.cm}
    \includegraphics[totalheight=5cm]{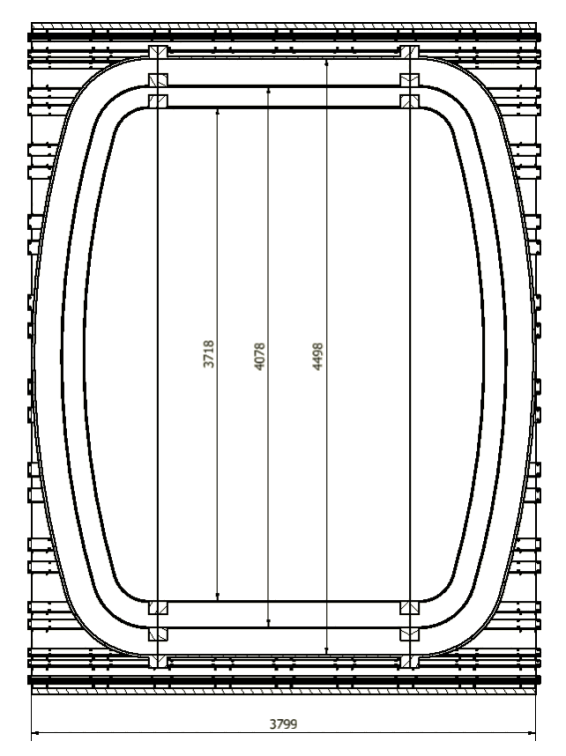}
    \caption{Left: 3D sketch of the cryostat inserted into the SC magnet, together with the 70 K screen and the copper cavity. Right: Preliminary design of cryostat and cavity.}
    \label{fig:cryostat}
  \end{center}
\end{figure}

In order to run at 4 K, the copper resonant-cavity will be hosted in a dedicated cryostat (Fig.~\ref{fig:cryostat}) composed of a vacuum chamber and a radiation shield at $\sim$ 70 K surrounding the cavity. The preliminary design foresees a cavity diameter of 3.72~m with maximum length of 3~m corresponding to a total volume of about 33~m$^3$. The total weight of cryostat and cavity is expected to be about 12 tons, driven by mechanical constraints.

The nominal resonance-frequency of the cavity is 64~MHz with quality factor about 750,000 at 4K assuming copper with RRR=25. Simulations show that a conventional system of two metallic tuning rods with 300~mm radius shift the resonance frequency from 70 to 110~MHz while keeping the quality factor above 500,000. Several strategies are under investigation to tune the frequency up to 250~MHz, including a multiple-cell
cavity~\cite{Pizzac} or its replacement with smaller ones, allowing us to probe the axion-mass region between 0.3 and 1~$\mu$eV. From preliminary calculations the quality factor is expected to scale almost linearly with frequency from 550,000 at 70 MHz down to 370,000 at 250 MHz. The cavity mode will be critically coupled to an electric-dipole field-probe and the signal read-out with a conventional or microstrip SQUID amplifier. A $^3$He refrigerator will cool down the SQUID amplifier to reduce its noise temperature to 300~mK~\cite{Squid}.

\section{The KLASH Sensitivity}\label{sec:sensitivity}
\begin{figure}[htbp]
  \begin{center}
    \includegraphics[totalheight=8cm]{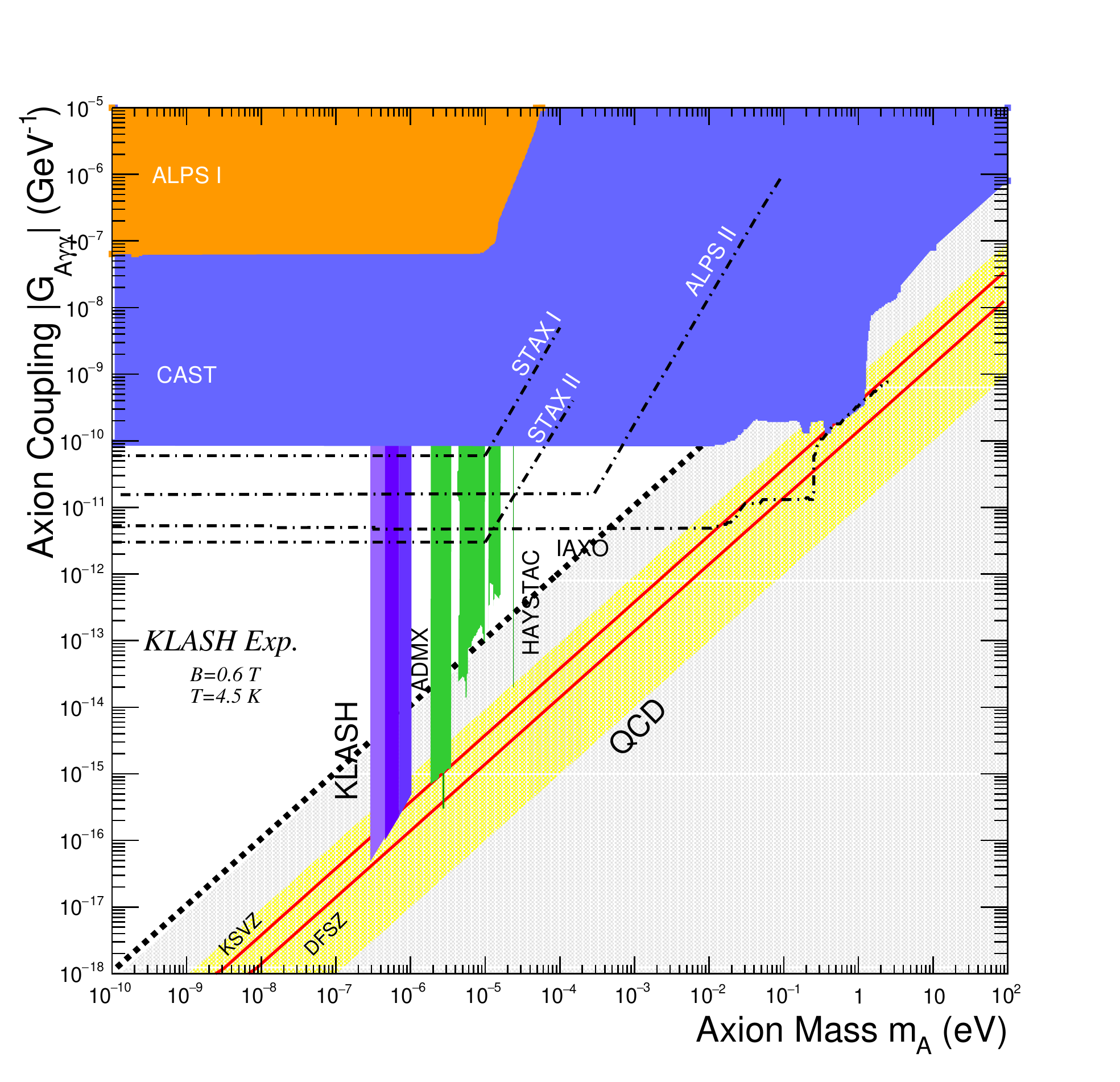}
    \caption{The KLASH discovery potential for axion mass $m_a$ in the range between 0.3 and 1 $\mu$eV, in three years of data taking.
    The three purple bands correspond to data taking with three different cavities of radii 1.9, 1.2 and 0.9~m.
    The yellow band is the preferred region for the KSVZ and DFSZ models. The gray region below the dashed line refers to the axion models discussed in~\cite{NARDI}. Also shown the expected sensitivity of STAX~\cite{STAX}, IAXO~\cite{IAXO} and ALPSII~\cite{ALPS2}.}
    \label{fig:sensitivity}
  \end{center}
\end{figure}

In this section, we assume to perform the frequency scan by employing three different cavities of radii 1.9, 1.2 and 0.9~m with two tuning rods both of radius 300, 190 and 140~mm, respectively. The integration time for each frequency step is expected to be between 10 and 15 minutes. The discovery potential in the coupling-mass plane, calculated with these assumptions, is shown in Figure~\ref{fig:sensitivity}. The sensitivity band reaches the region predicted for QCD axions of the KSVZ and DFSZ models in the mass range between 0.3 and 1~$\mu$eV.
The three vertical purple-bands correspond to one year of data taking with each of the three resonant cavities.

\section{Conclusions}
At the moment of writing, INFN funded a design study of one year for the KLASH Conceptual Design Report. One of the main outcomes of this study will be the mechanical design of the cryostat that will allow us to define the cost of construction.


\end{document}